\documentclass[preprint,showpacs,showkeys,preprintnumbers,amsmath,amssymb]{revtex4}
\usepackage{graphicx,caption2,subfigure,dcolumn}
\begin{document}

\title{Rapidity Spectra of the Secondaries Produced in Heavy Ion Collisions
and the Constituent Picture of the Particles}

\author{Bhaskar De}
 \email{bhaskar@imsc.res.in}
 \affiliation{Institute of Mathematical Sciences,\\
C.I.T. campus, Taramani, Chennai-600113, India}

\author{S. Bhattacharyya\footnote{Communicating Author}}
 \email{bsubrata@isical.ac.in}
 \affiliation{Physics
and Applied Mathematics Unit(PAMU),\\
Indian Statistical Institute, Kolkata - 700108, India}

\date{\today}

\begin{abstract}
The present study is both a reanalysis and an extension of the
approach initiated by Eremin and Voloshin(referenced in the text).
We attempt to interpret here the rapidity-spectra of the various
particles produced in both $Pb+Pb$ and $Au+Au$ collisions at
CERN-SPS and RHIC-BNL. The study made here is wider in scope and
is more species-specific than the earlier ones for which the results
obtained here have been compared with those suggested by 
some previous works based on HIJING, VENUS etc, at various
centralities. The study reconfirms that the constituent parton picture of 
the particles provides a better and more unified description of the
rapidity-density yields for the various secondaries, including some light cluster 
particles like deuteron even in heavy ion collisions.
\end{abstract}


\maketitle

\newpage
\section{Introduction}
In the recent past Eremin and Voloshin\cite{Eremin1} proposed
that both nucleons and individual nuclei could be considered as
superposition of the constituents (of the particles) called
partons/quarks/valons etc. Normally it is assumed that the 
nucleons are built of three such constituent
partons and the mesons are composed of two of them. Such a
constituent picture helps us to interpret the deep inelastic
collision processes involving lepton-hadron, hadron-hadron and
hadron-nucleus collisions. 
Much earlier, Bialas et al studied to explore the hypothesis,
within the framework of the additive quark model(AQM)\cite{Bialas1}, that the
constituent partons(quarks) provide the universal elements not only 
of nucleon-nucleon and nucleon-nucleus reactions but also nucleus-
nucleus interactions at high energies; this study ended up with surely
an affirmative indication to such a possibility.But, as the focus was on 
different issues,the central theme of this work deviated from what it is 
intended to be here.We will simply look into the behaviour of some
chosen observables studied here for 
nucleus-nucleus collisions alone from the viewpoint of such partonic constituent
pictures. The centrality-dependence of
multiplicity-density offers a very fundamental observable and we
present here our studies on production of pions, kaons, protons
and antiprotons separately in both $Pb+Pb$(including deuteron production for 
this collision) at CERN-SPS and $Au+Au$
collisions at RHIC-BNL at $\sqrt{s_{NN}}=130$ and 200 GeV respectively.
Besides, very recently the data on deuteron production in $Pb+Pb$ collision have 
been obtained, for which they have also been taken here 
into account. This apart,we will also compare the 
performance of the present approach with those obtained
on the basis of some standard-version programmings like
HIJING, VENUS etc which are essentially much more
complicated than the one dealt with here. Our
objective here is to examine the efficacy of this approach in explaining some
relevant data on production of these very important kind of
secondaries in the few now available high energy nuclear
collisions. The approach was also successfully
tested for production of neutral particles, like photons, by
Netrakanti and Mohanty\cite{Netrakanti1} in the very recent past
and was found to be in accord with the contention of Eremin and
Voloshin\cite{Eremin1}.

\section{The Outline of the Calculation of the Number of Participants}
In course of our calculations for the number of participant
nucleons, denoted by $N_{n-part}$, and the number of participant
partons, denoted by $N_{q-part}$ we follow exactly the ways
adopted by Eremin and Voloshin\cite{Eremin1} and Netrakanti and
Mohanty\cite{Netrakanti1}, wherein a Woods-Saxon nuclear density
profile is taken into account. This is given by

\begin{equation}
n_A(r) ~ = ~ \frac{n_0}{1+\exp{[(r-R)/d]}}
\end{equation}
where $n_0=0.17 ~ fm^{-3}$, $R=(1.12 A^{1/3} - 0.86 A^{-1/3}) ~
fm$, $d=0.54 ~ fm$.
\par
The number of participant nucleons($N_{n-part}$) for a nucleus-nucleus($A+B$) 
collision at an impact parameter $b$ is
calculated using the expression\cite{Eremin1,Netrakanti1},

\begin{equation}
\begin{array}{lcl}
N_{n-part}|_{AB} ~ & = & ~ \int ~ d^2s ~ T_A(\vec{s}) ~
\{1 ~ - ~ [ ~ 1 ~ - ~ \frac{\sigma_{NN}^{inel} T_B(\vec{s}-\vec{b})}{B} ~ ]^B\}\\
& + & ~ \int ~ d^2s ~ T_B(\vec{s}-\vec{b}) ~ \{~ 1 ~ -[~ 1 ~ - ~
\frac{\sigma_{NN}^{inel} T_A(\vec{s})}{A} ~ ]^A\},
\end{array}
\end{equation}

where $A$ and $B$ are the mass numbers of the two colliding
nuclei, $T(b) =  \int_{-\infty}^\infty  dz  n_A(\sqrt{b^2+z^2})$
is the thickness function, and $\sigma_{NN}^{inel}$ is the inelastic
nucleon-nucleon cross section. The
number of participant partons is also calculated in a similar
manner by taking into account the following changes of the facts
related to physical realities, viz., (i) the density is changed
to three times that of nucleon density with $n_0^q=3n_0=0.51
fm^{-3}$; (ii) the cross sections are changed to
$\sigma_{qq}=\sigma_{NN}^{inel}/9$ and (iii) the mass numbers of the
colliding nuclei are changed to three times of their values keeping the 
size of the nuclei same as in the cases of $N_{n-part}$\cite{Eremin1,Netrakanti1}.
\par
In the same way, the number of participant partons in a $PP$ or $P\bar{P}$ collision
can also be calculated by using $A=3$ and $B=3$, and considering nucleons as hard 
spheres of uniform radii 0.8 $fm$\cite{Wong1}.
\par
For the present work we use
$\sigma_{NN}^{inel} =$ 30 mb in the range of c.m. energies $\sqrt{s_{NN}}=$ 4-17.3
GeV as suggested in Ref.\cite{Misko1}. For higher energies, we first fit the data on
total cross-sections for $PP$ and $P\bar{P}$ collisions\cite{PDG} at energies
beyond $\sqrt{s_{NN}}= 30$ GeV
with the expression\cite{De1}
\begin{equation}
\sigma_{NN}^{total} ~ = ~ (29.2\pm0.3)[1+\frac{(2.2\pm0.5) ~ GeV}{\sqrt{s_{NN}}}]+
(1.1\pm0.1)\ln{(\frac{s_{NN}}{10 ~ GeV^2})} + (0.19\pm0.01)\ln^2{(\frac{s_{NN}}{10 ~
GeV^2})} 
\end{equation}

with $\chi^2/ndf=171/47$, where $ndf=$ No. of data $-$ No. of Parameters. We also fit
the elastic cross-section data for the same interactions\cite{PDG} with a similar type of
expression which is given by,

\begin{equation}
\sigma_{NN}^{elastic} ~ = ~ (6.2\pm0.2)[1+\frac{(4.8\pm1.4) ~ GeV}{\sqrt{s_{NN}}}]-
(0.47\pm0.04)\ln{(\frac{s_{NN}}{10 ~ GeV^2})} + (0.11\pm0.01)\ln^2{(\frac{s_{NN}}{10 ~
GeV^2})} 
\end{equation}

with $\chi^2/ndf=53.14/18$. The fits along with the data are shown in Fig.1. 
Hence, one can obtain the energy-dependence of inelastic 
cross-section($\sigma_{NN}^{inel}$) for nucleon-nucleon interaction by 
subtracting eqn.(4) from eqn.(3). The values of $\sigma_{NN}^{inel}$, obtained in this
manner, at some energies are given in Table-\ref{tab:table1}. The obtained values fall within $6\%$ error of 
those suggested in Ref.\cite{Misko1}. But a point is to be
noted. Though some sort of justification for proposing the nature     
of the total cross section in the form of expression (3) given in Ref.\cite{De1},
none such explanation for expression (4) is possible. This is simply 
assumed with a view to providing only a best fit to the available data on 
elastic cross section shown in Fig.1 and extracting a workable phenomenological form
of expression for the nature of inelastic cross section, $\sigma_{NN}^{inel}$.
In fact, these twin relations help us to arrive at the usable
values of inelastic cross sections needed for our necessary calculations
of the relevant observables in a systematic manner even at energies
for which no  measured data on cross section values are available.

\section{Results and Discussions}

The values of the number of participant nucleons and those of the 
participant partons for three nucleus-nucleus collisions at various 
energies and different centralities are presented in the Tables(\ref{tab:table2}-
\ref{tab:table4}). A comparison of the results obtained on the basis of the 
present work with those of NA49\cite{Bachler1}, WA98\cite{Aggarwal1} and 
PHENIX\cite{Adcox1,Adcox2,Adler1}
groups is made in Fig.-2(a) to Fig.-2(d), in terms of 
$1-\frac{<N^{Present Work}_{n-part}>}{<N^{NA49/WA98/PHENIX}_{n-part}>}$ 
versus $N^{NA49/WA98/PHENIX}_{n-part}$ . It is observed that 
the agreements in cases of Au+Au collisions are  modestly fair,
whereas for Pb+Pb interaction  the disagreement is quite strong and 
prominent. This discrepancy could, for the present, be attributed to
the much less c.m energy in lead-lead reaction which falls roughly at
17.2 GeV only. 

\par
The integrated yields of the identified hadrons at central rapidity region produced
in $Pb+Pb$ and $Au+Au$ collisions at different centralities are presented 
diagrammatically in Fig.3-Fig.5. In subfigures labelled as (a) and (b) of each category, 
the experimental data on integrated yields are normalized by half of the number of 
participant nucleons($N_{n-part}$) while those in subfigures labelled as (c) and (d) are normalized
by half of the number of participant partons($N_{q-part}$). As we are here to make
comparison between these two cases, we should use values of both the observables
obtained from a single model. And that is why we put into use the values of 
$<N_{n-part}>$ obtained on the basis of the present work,instead of the values 
indicated by NA49\cite{Bachler1} or PHENIX groups\cite{Adcox1,Adcox2,Adler1}.
\par
Now, to make a comparison between these two cases we fit the data in Fig.3-Fig.5 
by two phenomenological expressions which are given by,

\begin{equation}
\frac{1}{0.5 ~ <N_{n-part}>}\frac{dN_{ch}}{dy} ~ = ~ a ~ N^{\alpha}_{n-part}
\end{equation}

and
 
\begin{equation}
\frac{1}{0.5 ~ <N_{q-part}>}\frac{dN_{ch}}{dy} ~ = ~ b ~ N^{\beta}_{q-part}
\end{equation}

where $a$, $b$, $\alpha$ and $\beta$ are four constants. The fitted values of these
parameters are given in Tables(\ref{tab:table5}-\ref{tab:table7}) and depicted by solid curves in the figures. 
As could be seen from Fig.2 that the most peripheral
values of $N_{n-part}$ for $Pb+Pb$ collision at $E_{Lab}=158A$ GeV and $Au+Au$
collision at $\sqrt{s_{NN}}=200$ GeV obtained by the present model show larger
discrepancies with respect to those obtained by NA49 and PHENIX collaborations, we keep
most peripheral data in these two collisions out of the range of both the fits provided
by eqn.(5) and eqn.(6).
\par
Obviously, the factors $\alpha$ and $\beta$ provide the slopes of the fits. As the
magnitude of any of these factors takes value nearer to zero, the fit will exhibit
better flatness of the data with respect to $N_{n-part}$ or $N_{q-part}$ implying a
superior compliance of scaling , i.e., 
a better scaling. A look at Table-\ref{tab:table5} reveals that the 
data on $<\pi>$, $K^{\pm}$ and $P/\bar{P}$ in $Pb+Pb$ collisions 
show a better degree of conformity
with scaling when  normalized by half of $N_{q-part}$, as in all cases
$|\beta|$ is less than $|\alpha|$. But,for the case of deuteron production in 
$Pb+Pb$ collision, the scenario is just the opposite. Besides, the 
integrated yields for charged pions produced in $Au+Au$ interactions
at both the RHIC energies[Tables~\ref{tab:table6} - \ref{tab:table7}] do not exhibit scaling satisfactorily,
when the data are normalized by half of $N_{q-part}$ instead of 
$N_{n-part}$, as the values of $|\alpha|$, in these cases, are quite
small with respect to $|\beta|$. However, the data on $K^{\pm}$ and $P/\bar{P}$
produced in the same collisions favour the $N_{q-part}$-scaling over
the $N_{n-part}$-scaling, though the cases of $P/\bar{P}$ at $\sqrt{s_{NN}}=200$ GeV 
are not as prominent as in the other cases. In fine, our net finding from the present
study is the pions do not agree, while the kaons and protons do. 
\par     
The Fig.6(a) deserves some special attention and comments. The
plots on integrated yields for charged hadrons produced in 
both nucleus-nucleus($A+A$) and $P+\bar{P}$ collision deflect
from each other when data on both are separately normalized in
terms of number of participating nucleons($N_{n-part}$). The data on integrated yields
in central nucleus-nucleus collisions normalized by the number of participant
nucleon-pair have been used here from Fig.3 of Ref.\cite{Back1}.
To our purpose,this has to be normalized in terms of the basic 
parton(quark)-constituents, denoted by $<N_{q-part}>$, 
for which the (charged)pseudo-
rapidity density terms for various nucleus-nucleus collisions
normalized by participant nucleon-pair, i.e., the factors
$\frac{1}{0.5<N_{n-part}>}\frac{dN_{ch}}{d\eta}$ are to be multiplied by 
$R=<N_{n-part}>/<N_{q-part}>$ with the values of them as given in Table-\ref{tab:table8}. 
And, in this conversion we have utilized those particular values of $<N_{n-part}>$
as were used in Ref.\cite{Back1}, so that we can normalize the exact values
of $\frac{dN_{ch}}{d\eta}$ by $<N_{q-part}>/2$. In calculating $N_{q-part}$ for
$P+\bar{P}$ collisions, we use equation(2) and the values of $N_{q-part}$ for
$P+\bar{P}$ collisions at three different c.m. energies are given in Table-\ref{tab:table9} as a
function of relative probability. 
\par
And when partonic considerations are used in normalization, 
the data on both $P+\bar{P}$ and $A+A$ 
collisions come to an agreeable state. 
The $<N_{q-part}>$-values that are to be used for 
normalization of the most central $P+\bar{P}$ collisions, i.e.,$0-5\%$
central collisions are depicted in Table-\ref{tab:table10}. In order to check the nature of agreement
between the data on $A+A$ and $P+\bar{P}$ collisions, 
we try to obtain a fit by taking into 
account both sets of data, normalized by $<N_{q-part}>/2$; and the desired fit is
to be described here by the following expression:

\begin{equation}
\frac{1}{0.5 ~ <N_{q-part}>}\frac{dN_{ch}}{d\eta} ~ = ~ -(0.010\pm 0.003) ~ + ~ (0.27\pm0.01) ~
\ln{(\frac{\sqrt{s_{NN}}}{GeV})}
\end{equation}

with $\chi^2/ndf = 3.632/12$. The goodness of the fit is also shown in Fig.6(b) with the 
help of the expression $fit-data$ as a function of $\sqrt{s_{NN}}$ which reveals that
both sets of data are in good agreement with respect to the fit. Hence, both $A+A$ and
$P+\bar{P}$ data exhibit a common $\sqrt{s_{NN}}$-dependence when the data on the 
former are normalized by the number of participant parton-pairs.   
Very recently, Sarkisyan and Shakarov\cite{Sarkisyan1} adopted a somewhat similar type approach in dealing with data on pseudorapidity density per participant-pair in cases of hadron-hadron, nucleus-nucleus collisions and also in the cases of $e^+e^-$ interactions. The study thus reinforced the idea of similarities of the bulk observables in the widest possible ranges of particle-interactions. 
\section{Concluding Remarks}

Let us now summarize our observations made here: (i) The
secondaries, excluding deuterons, produced in $Pb+Pb$ collision at CERN-SPS behave
modestly well vis-a-vis $N_{q-part}$ scaling. The behaviour is
more consistent towards the highest values of centrality, with
centrality maximum at 0-5$\%$(represented by solid boxes in the
graph). (ii) Similar statements could be made about $Au+Au$
interactions at both energies(Fig.4 and Fig.5) except the cases for pions. (iii)
It is quite noticeable that there are modest degree of divergences
in the $N_{n-part}$-values between the calculations done by us at different
centralities and those by others, as indicated in the text (in the 
second and third columns of the various Tables presented in this 
work). For the present, these discrepancies could be attributed only to 
our much simpler programming than what is resorted to by the big 
groups like NA49, WA98, PHENIX etc. The plain fact is we are now simply
unable to take up such rigorous studies as might be desired or 
advisable due to various reasons beyond our capacity and control.
But, we recognize the urgency and importance of such studies in
order to arrive at a decision about the merit of $N_{q-part}$ scaling
and of the viewpoints expressed by Eremin and Voloshin\cite{Eremin1}.
(iv) The Fig.6 highlights on how the data on both $P+\bar{P}$ and $A+A$ collisions
negotiate better the idea of partonic-participant scaling, when the
constituent-participant number for both $P+\bar{P}$ and $A+A$ collisions are
taken into account. Thus, the present 
study does essentially provide modest support to the viewpoints expressed 
by Eremin and Voloshin in their work of Ref.\cite{Eremin1}.

\begin{acknowledgments}
The authors are very grateful to the anonymous referee for his/her
very patient readings, constructive
criticisms and kind valuable instructions for the improvements in the
previous drafts of the manuscript.
The authors would also like to express their thankful gratitude to Professor D.
Mi$\grave{\rm{s}}$kowiec for providing some helpful suggestions
about running the FORTRAN code developed by him in calculating the
number of participants, $N_{part}$. One of the authors,BD, is thankful
to the IMSc for offering him the support by a post-doctoral Fellowsip 
of the IMSc, wherein a part of this work was done.
\end{acknowledgments}


\newpage
\begin{table}
\caption{\label{tab:table1}Values of inelasitc nucleon-nucleon cross-section at different energies. The
second column provides the values obtained by subtracting eqn.(4) from eqn.(3). The
last column gives the values as cited in Ref.\cite{Misko1}.}
\begin{ruledtabular}
\begin{tabular}{ccc}
\hline $\sqrt{s_{NN}} (GeV)$ & $\sigma_{NN}^{inel}$(mb) &
$\sigma_{NN}^{inel}$(mb)\\
& (Present Work) & (From Ref.\cite{Misko1})\\
\hline
53 & 35 & -\\
56 & 35 & 37\\
130 & 40 & 41\\
200 & 42 & 42\\
540 & 48 & -\\
630 & 49 & -\\
900 & 51 & -\\
1800 & 56 & -\\
\hline
\end{tabular}
\end{ruledtabular}
\end{table}

\begin{table}
\caption{\label{tab:table2}Values of $<N_{n-part}>$ and $<N_{q-part}>$ for $Pb+Pb$ collisions at
$E_{Lab}=158A$ GeV. }
\begin{ruledtabular}
\begin{tabular}{cccc}
\hline Centrality & $<N_{n-part}^{NA49}>$\cite{Bachler1} & $<N_{n-part}^{Present Work}>$
& $<N_{q-part}^{Present Work}>$\\
\hline
$0-5\%$ & 362 & 368 & 856\\
$5-14\%$ & 305 & 290 & 644\\
$14-23\%$ & 242 & 210 & 446\\
$23-32\%$ & 189 & 140 & 278\\
$32-47\%$ & 130 &  92 & 171\\
$47-99\%$ & 72 & 32 & 53\\
\hline
\hline Centrality & $<N_{n-part}^{WA98}>$\cite{Aggarwal1} & $<N_{n-part}^{Present Work}>$
& $<N_{q-part}^{Present Work}>$\\
\hline
$0-1\%$ & $380\pm1$ & 385 & 901\\
$1-6.8\%$ & $346\pm1$ & 338 & 770\\
$6.8-13\%$ & $290\pm2$ & 273 & 600\\
$13-25.3\%$ & $224\pm1$ & 196 & 411\\
$25.3-48.8\%$ & $132\pm3$ &  99 & 189\\
$48.8-67\%$ & $63\pm2$ & 35 & 56\\
$67-82.8\%$ & $30\pm2$ & 10 & 14\\
$82.8-100\%$ & $12\pm2$ & 3 & 4\\
\hline
\end{tabular}
\end{ruledtabular}
\end{table}

\begin{table}
\caption{\label{tab:table3}Values of $<N_{n-part}>$ and $<N_{q-part}>$ for $Au+Au$ collisions at 
$\sqrt{s_{NN}}=130$ GeV.}
\begin{ruledtabular}
\begin{tabular}{cccc}
\hline Centrality & $<N_{n-part}^{PHENIX}>$\cite{Adcox1,Adcox2} & $<N_{n-part}^{Present Work}>$
& $<N_{q-part}^{Present Work}>$ \\
\hline
$0-5\%$ & $347.7\pm10$ & 351 & 880\\
$5-10\%$ & $293\pm10$ & 297 & 710.6\\
$5-15\%$ & $271.3\pm8.4$ & 272 & 645\\
$10-15\%$ & $248\pm8$ & 242.5 & 565.4\\
$15-20\%$ & $211\pm7$ & 207 & 469.4\\
$20-25\%$ & $177\pm7$ & 173 & 381\\
$15-30\%$ & $180.2\pm6.6$ & 174.5 & 387.3\\
$25-30\%$ & $146\pm6$ & 139 & 298\\
$30-35\%$ & $122\pm5$ & 120 & 249\\
$35-40\%$ & $99\pm5$ & 96 & 192\\
$40-45\%$ & $82\pm5$ & 81.5 & 158.7\\
$45-50\%$ & $68\pm4$ & 62 & 114.6\\
$30-60\%$ & $78.5\pm4.6$ & 77 & 150.1\\
$60-92\%$ & $14.3\pm3.3$ & 11.8 & 17.8\\
\hline
\end{tabular}
\end{ruledtabular}
\end{table}

\begin{table}
\caption{\label{tab:table4}Values of $<N_{n-part}>$ and $<N_{q-part}>$ for $Au+Au$ collisions at 
$\sqrt{s_{NN}}=200$ GeV.}
\begin{ruledtabular}
\begin{tabular}{cccc}
\hline Centrality & $<N_{n-part}^{PHENIX}>$\cite{Adler1} & $<N_{n-part}^{Present Work}>$ & $<N_{q-part}^{Present Work}>$\\
\hline
$0-5\%$ & $351.4\pm2.9$ & 351.9 & 880.1\\
$0-10\%$ & $325.2\pm3.3$ & 332.5 & 819.7\\
$5-10\%$ & $299.0\pm3.8$ & 292.2 & 694.6\\
$10-15\%$ & $253.9\pm4.3$ & 242.4 & 560.9\\
$10-20\%$ & $234.6\pm4.7$ & 226.2 & 517.4\\
$15-20\%$ & $215.3\pm5.3$ & 208.2 & 469\\
$20-30\%$ & $166.6\pm5.4$ & 157 & 339\\
$30-40\%$ & $114.2\pm4.4$ & 108.4 & 218\\
$40-50\%$ & $74.4\pm3.8$ & 71.5 & 134.3\\
$50-60\%$ & $45.5\pm3.3$ & 40.4 & 68.8\\
$60-70\%$ & $25.7\pm3.8$ & 23 & 35.6\\
$60-80\%$ & $19.5\pm3.3$ & 16.6 & 25\\
$60-92\%$ & $14.5\pm2.5$ & 11.8 & 17.3\\
$70-80\%$ & $13.4\pm3.0$ & 9.9 & 13.6\\
$70-92\%$ & $9.5\pm1.9$ & 6.8 & 9.1\\
$80-92\%$ & $6.3\pm1.2$ & 3.9 & 5\\
\hline
\end{tabular}
\end{ruledtabular}
\end{table}

\begin{table}
\caption{\label{tab:table5}Values of $a$, $b$, $\alpha$ and $\beta$ for production of various secondaries
in $Pb+Pb$ collisions at $E_{Lab}=158A$ GeV[ndf=No. of data - No. of parameters in the
fit].}
\begin{ruledtabular}
\begin{tabular}{ccccccc}
\hline Identified & $a$ & $\alpha$ & $\chi^2/ndf$ & $b$ & $\beta$ & $\chi^2/ndf$\\
Secondary & &  & & & & \\
\hline
$<\pi>$ & $0.74\pm0.06$ & $0.13\pm0.01$ & $0.320/3$ & $0.82\pm0.08$ & $-(0.02\pm0.01)$
& $0.442/3$\\
$K^+$ & $0.023\pm0.005$ & $0.42\pm0.04$ & $3.436/3$ & $0.026\pm0.006$ & $0.22\pm0.04$
& $3.700/3$\\
$K^-$ & $0.028\pm0.008$ & $0.37\pm0.05$ & $11.024/3$ & $0.03\pm0.01$ &
$0.18\pm0.05$ & $12.514/3$\\
$P$ & $0.06\pm0.01$ & $0.17\pm0.04$ & $4.492/3$ & $0.06\pm0.01$ &
$0.03\pm0.01$ & $4.054/3$\\
$\bar{P}$ & $0.006\pm0.002$ & $0.24\pm0.06$ & $4.616/3$ & $0.007\pm0.002$ &
$0.07\pm0.04$ & $4.313/3$\\
$D$ & $0.0015\pm0.0004$ & $0.05\pm0.02$ & $0.635/3$ & $0.0017\pm0.0004$ &
$-(0.11\pm0.04)$ & $0.656/3$\\
\hline
\end{tabular}
\end{ruledtabular}
\end{table}
\begin{table}
\caption{\label{tab:table6}Values of $a$, $b$, $\alpha$ and $\beta$ for production of various secondaries
in $Au+Au$ collisions at $\sqrt{s_{NN}}=130$ GeV[ndf=No. of data - No. of parameters in the
fit].}
\begin{ruledtabular}
\begin{tabular}{ccccccc}
\hline Identified & $a$ & $\alpha$ & $\chi^2/ndf$ & $b$ & $\beta$ & $\chi^2/ndf$\\
Secondary & &  & & & & \\
\hline
$\pi^+$ & $1.6\pm0.1$ & $0.0015\pm0.0003$ & $0.250/3$ & $1.6\pm0.1$ & $-(0.13\pm0.01)$
& $0.221/3$\\
$\pi^-$ & $1.3\pm0.1$ & $0.02\pm0.01$ & $0.190/3$ & $1.32\pm0.07$ & $-(0.12\pm0.01)$
& $0.125/3$\\
$K^+$ & $0.11\pm0.01$ & $0.15\pm0.01$ & $0.073/3$ & $0.12\pm0.01$ &
$-(0.012\pm0.002)$ & $0.083/3$\\
$K^-$ & $0.11\pm0.03$ & $0.11\pm0.06$ & $1.175/3$ & $0.11\pm0.03$ &
$-(0.03\pm0.01)$ & $1.081/3$\\
$P$ & $0.09\pm0.01$ & $0.10\pm0.02$ & $0.191/3$ & $0.09\pm0.01$ &
$-(0.05\pm0.01)$ & $0.151/3$\\
$\bar{P}$ & $0.06\pm0.01$ & $0.10\pm0.02$ & $0.148/3$ & $0.06\pm0.01$ &
$-(0.04\pm0.01)$ & $0.115/3$\\
\hline
\end{tabular}
\end{ruledtabular}
\end{table}
\begin{table}
\caption{\label{tab:table7}Values of $a$, $b$, $\alpha$ and $\beta$ for production of various secondaries
in $Au+Au$ collisions at $\sqrt{s_{NN}}=200$ GeV[ndf=No. of data - No. of parameters in the
fit].}
\begin{ruledtabular}
\begin{tabular}{ccccccc}
\hline Identified & $a$ & $\alpha$ & $\chi^2/ndf$ & $b$ & $\beta$ & $\chi^2/ndf$\\
Secondary & &  & & & & \\
\hline
$\pi^+$ & $1.7\pm0.1$ & $-(0.008\pm0.004)$ & $4.455/8$ & $1.6\pm0.1$ & $-(0.12\pm0.01)$
& $0.769/8$\\
$\pi^-$ & $1.4\pm0.1$ & $0.03\pm0.01$ & $1.071/8$ & $1.5\pm0.1$ & $-(0.12\pm0.01)$
& $1.015/8$\\
$K^+$ & $0.130\pm0.004$ & $0.13\pm0.01$ & $0.583/8$ & $0.14\pm0.01$ &
$-(0.03\pm0.01)$ & $0.591/8$\\
$K^-$ & $0.12\pm0.01$ & $0.14\pm0.01$ & $1.657/8$ &
$0.13\pm0.01$ & $-(0.02\pm0.01)$ & $1.490/8$\\
$P$ & $0.063\pm0.002$ & $0.09\pm0.01$ & $0.264/8$ & $0.069\pm0.003$ &
$-(0.07\pm0.01)$ & $0.274/8$\\
$\bar{P}$ & $0.048\pm0.002$ & $0.08\pm0.01$ & $0.290/8$ & $0.053\pm0.002$ &
$-(0.07\pm0.01)$ & $0.347/8$\\
\hline
\end{tabular}
\end{ruledtabular}
\end{table}

\begin{table}
\caption{\label{tab:table8}Values of $<N_{n-part}>$ and $<N_{q-part}>$ used in Fig.6(a) to normalize 
$\frac{dN_{ch}}{d\eta}$-data for various nucleus-nucleus collisions} 
\begin{ruledtabular}
\begin{tabular}{ccccc}
\hline Collision type & $\sqrt{s_{NN}}$ (GeV) & Centrality & $<N_{n-part}>$ &
$<N_{q-part}>$\\
\hline
$Au+Au$ & AGS & $0-5\%$ & 343\cite{Klay1} & 786\\
$Pb+Pb$ & 8.7 & $0-5\%$ & 349\cite{Afanasiev1} & 856\\
$Pb+Pb$ & 17.2 & $0-5\%$ & 362\cite{Afanasiev1} & 856\\
$Au+Au$ & 56 & $0-6\%$ & 330\cite{Back2} & 823\\
$Au+Au$ & 130 & $0-6\%$ & 343\cite{Back2} & 871\\
$Au+Au$ & 200 & $0-6\%$ & 344\cite{Back1} & 871\\
\hline
\end{tabular}
\end{ruledtabular}
\end{table}

\begin{table}
\caption{\label{tab:table9}Values of $N_{q-part}$ for proton-antiproton collisions at three different c.m. energies
as a function of relative probability} 
\begin{ruledtabular}
\begin{tabular}{cccc}
\hline Relative Probability & 53 GeV & 200 GeV & 1800 GeV\\
\hline
0-5$\%$ & 3.2 & 3.5 & 4.0\\
5-10$\%$ & 2.8 & 3.1 & 3.7\\ 
10-20$\%$ & 2.5 & 2.8 & 3.3\\
20-30$\%$ & 2.1 & 2.3 & 2.7\\
30-40$\%$ & 1.9 & 2.0 & 2.5\\
40-60$\%$ & 1.4 & 1.5 & 1.7\\
60-100$\%$ & 0.9 & 0.9 & 1.0\\
\hline
\end{tabular}
\end{ruledtabular}
\end{table}

\begin{table}
\caption{\label{tab:table10}Values of $<N_{q-part}>$ for most central($0-5\%$) $P+\bar{P}$ collisions at different c.m. energies} 
\begin{ruledtabular}
\begin{tabular}{ccccccc}
\hline  $\sqrt{s_{NN}}$ (GeV) & 53  & 200 & 540 & 630 & 900 & 1800 \\
$<N_{q-part}>$ & 3.2 & 3.5 & 3.6 & 3.7 & 3.8 & 4.0 \\
\hline
\end{tabular}
\end{ruledtabular}
\end{table}

\newpage
\begin{figure}
   \centering
\includegraphics[width=8cm]{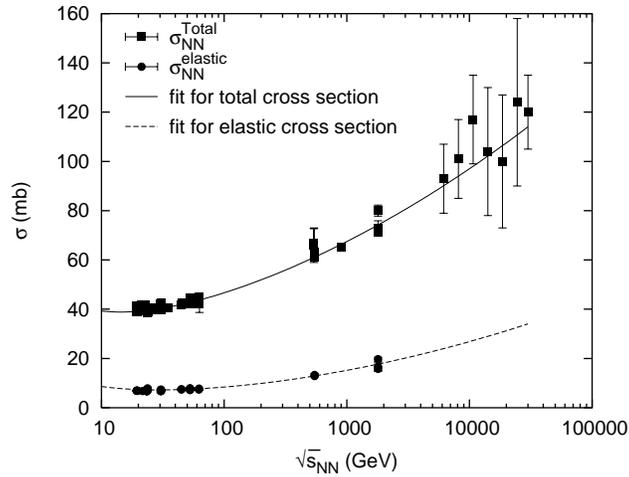}
\caption{Plots of total and elastic cross-section of $PP$ and $P\bar{P}$ interactions
as a function of c.m. energies.
Data points are taken from Ref.\cite{PDG}. The solid curve depicts the fit for total
cross-sections on the basis of eqn.(3) while the dashed one is for elastic
cross-section on the basis of eqn.(4).}
\end{figure}
 \begin{figure}
 \subfigure[]{
  \begin{minipage}{.5\textwidth}
   \centering
\includegraphics[width=8cm]{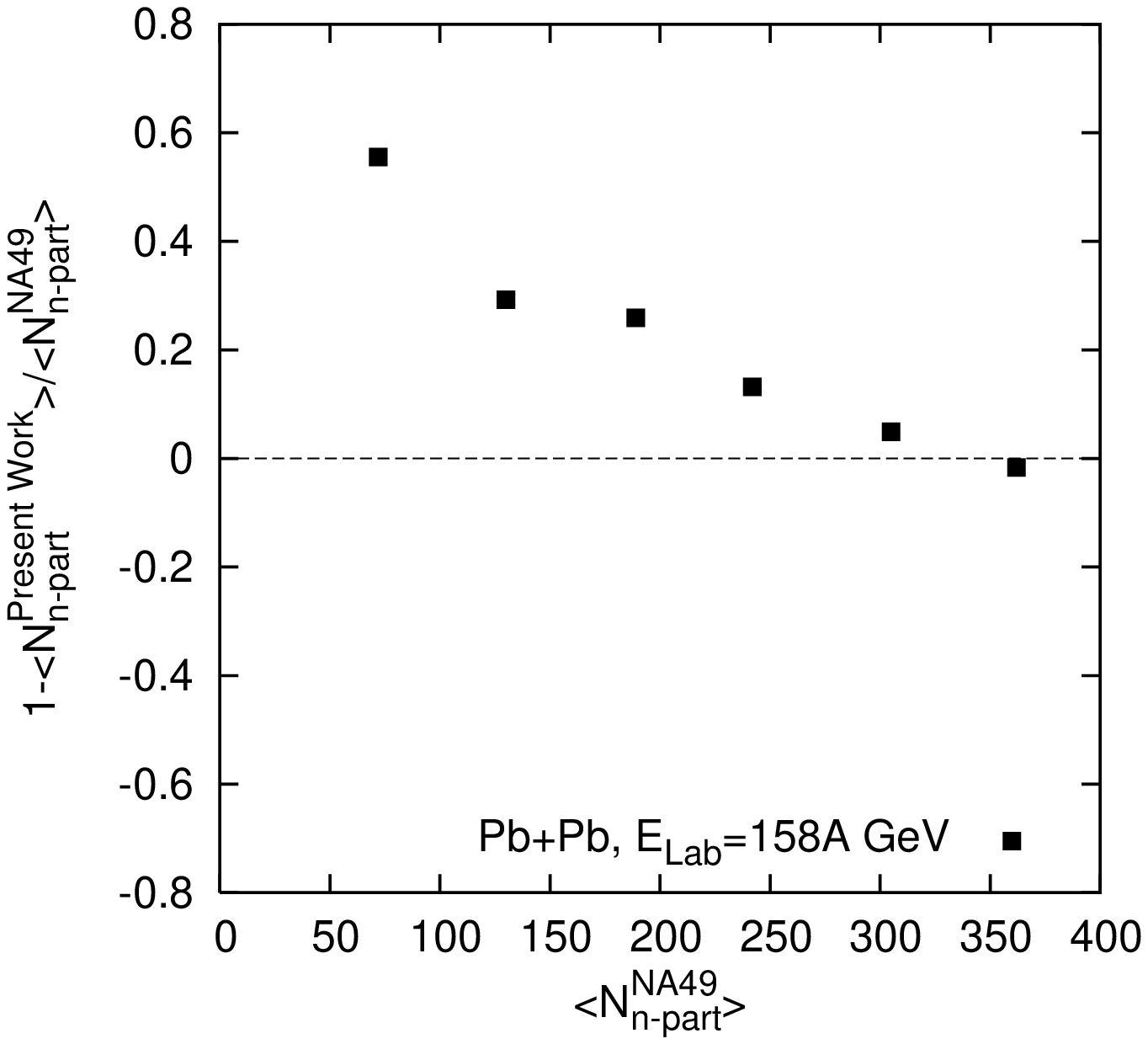}
  \end{minipage}}%
 \subfigure[]{
  \begin{minipage}{.5\textwidth}
   \centering
\includegraphics[width=8cm]{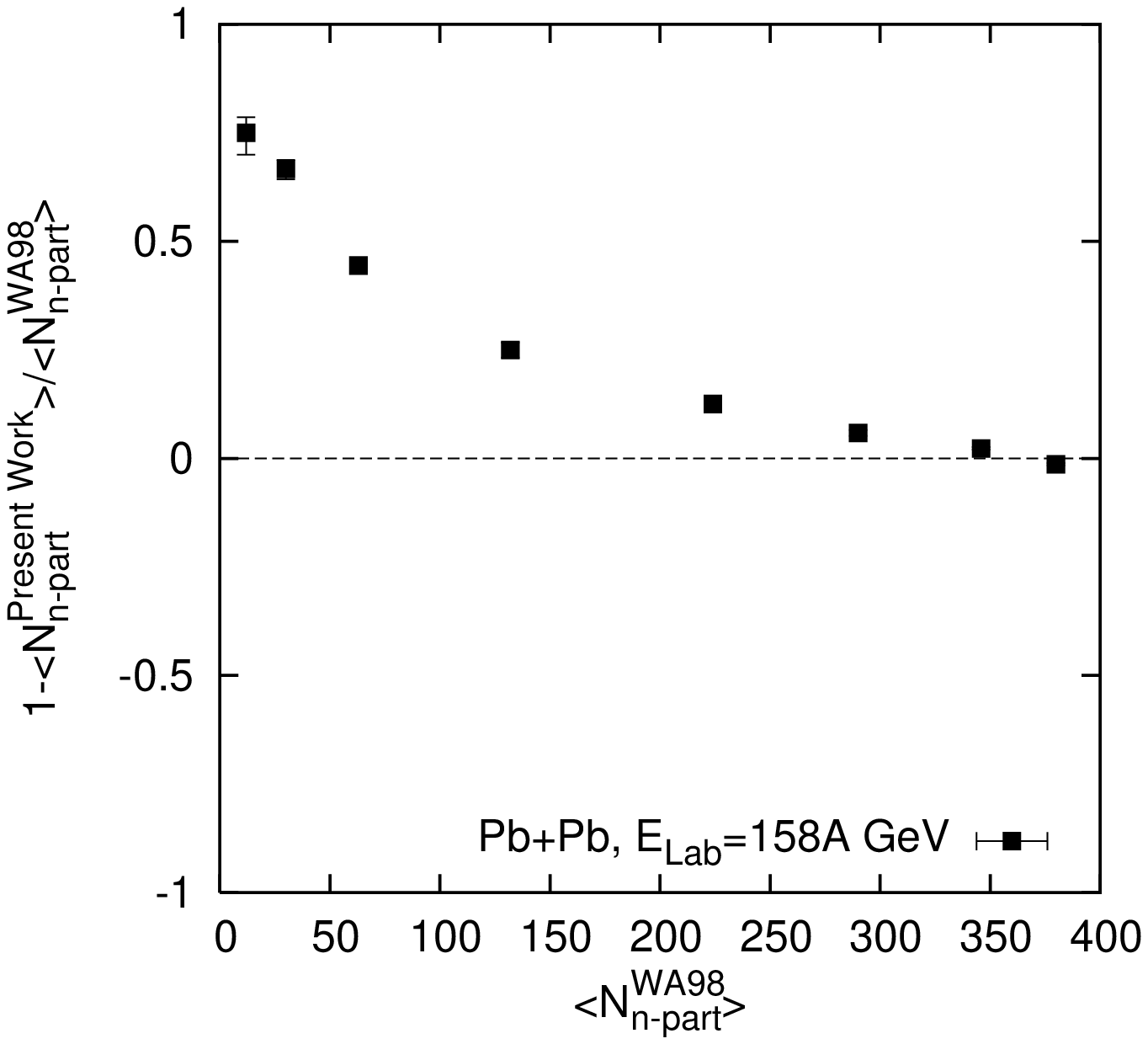}
  \end{minipage}}%
\vspace{.01cm}
 \subfigure[]{
  \begin{minipage}{.5\textwidth}
   \centering
\includegraphics[width=8cm]{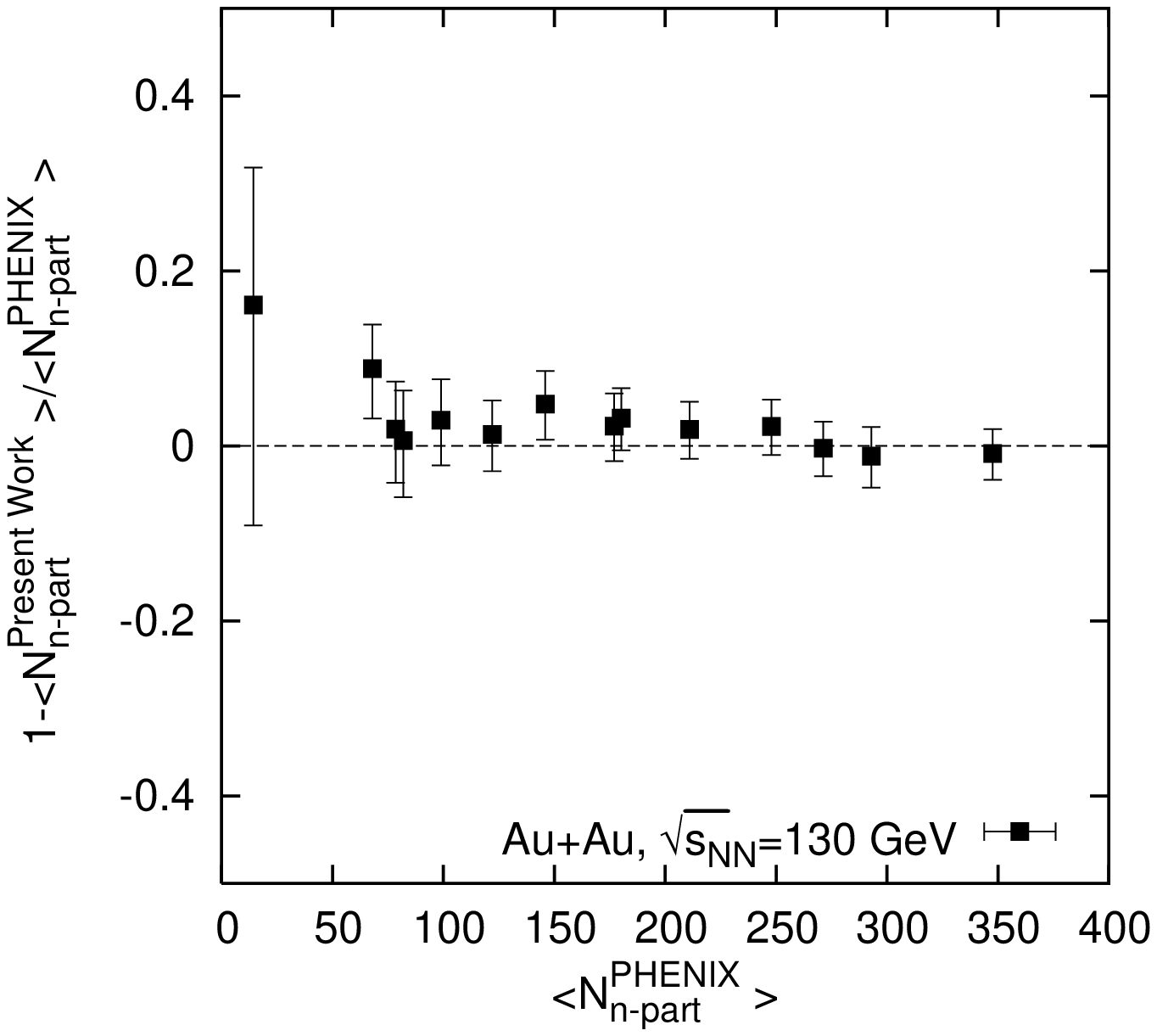}
  \end{minipage}}%
 \subfigure[]{
  \begin{minipage}{.5\textwidth}
   \centering
\includegraphics[width=8cm]{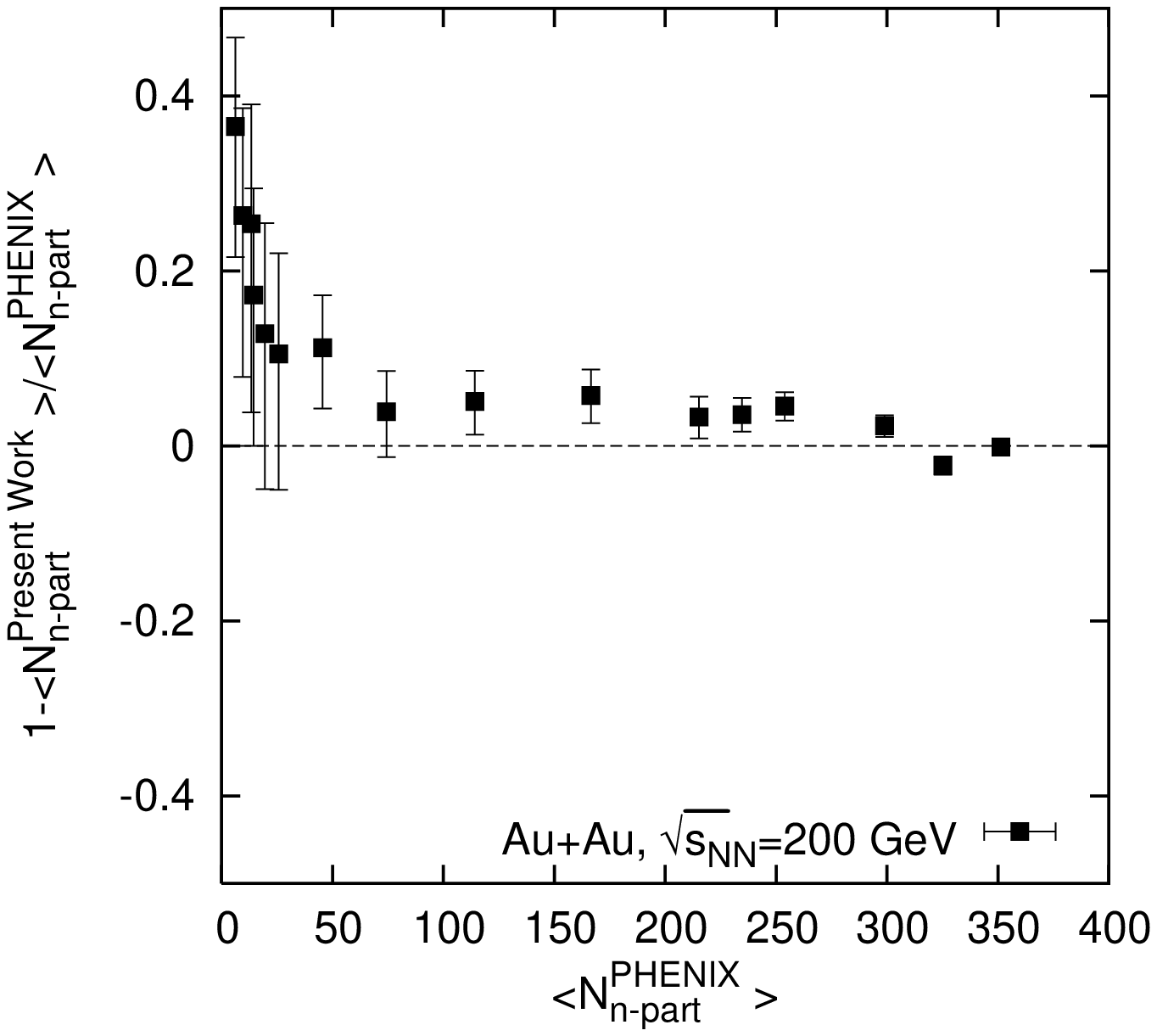}
  \end{minipage}}%
   \caption{Comparison of average number of participant nucleons obtained in the
   present work and by different experimental collaborations in various heavy ion
   collisions. The experimental data are taken from
   Ref.\cite{Adcox1,Bachler1,Aggarwal1,Adcox2,Adler1}.}
\end{figure}

\begin{figure}
 \subfigure[]{
  \begin{minipage}{.5\textwidth}
   \centering
\includegraphics[width=8cm]{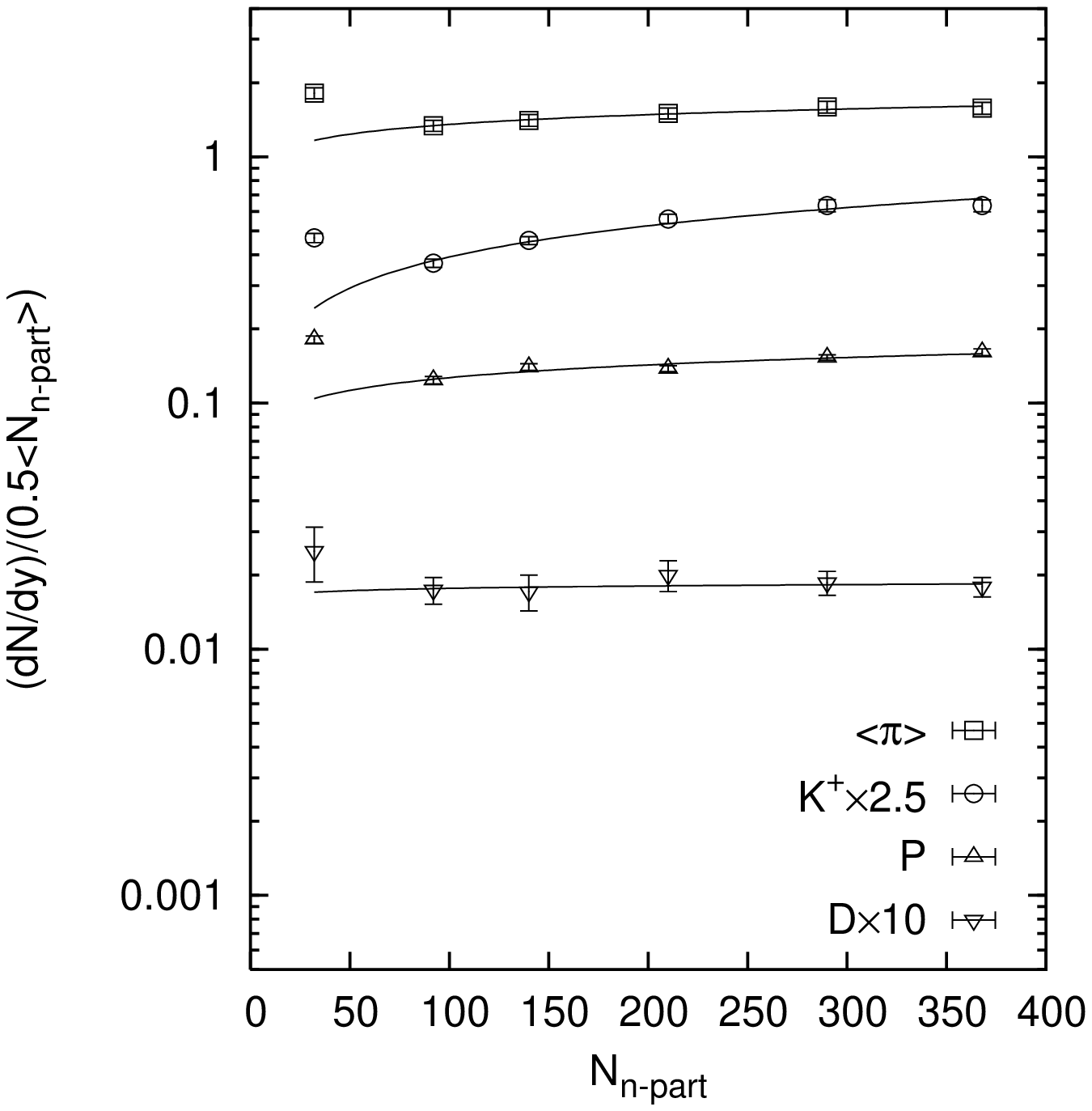}
  \end{minipage}}%
 \subfigure[]{
  \begin{minipage}{.5\textwidth}
   \centering
\includegraphics[width=8cm]{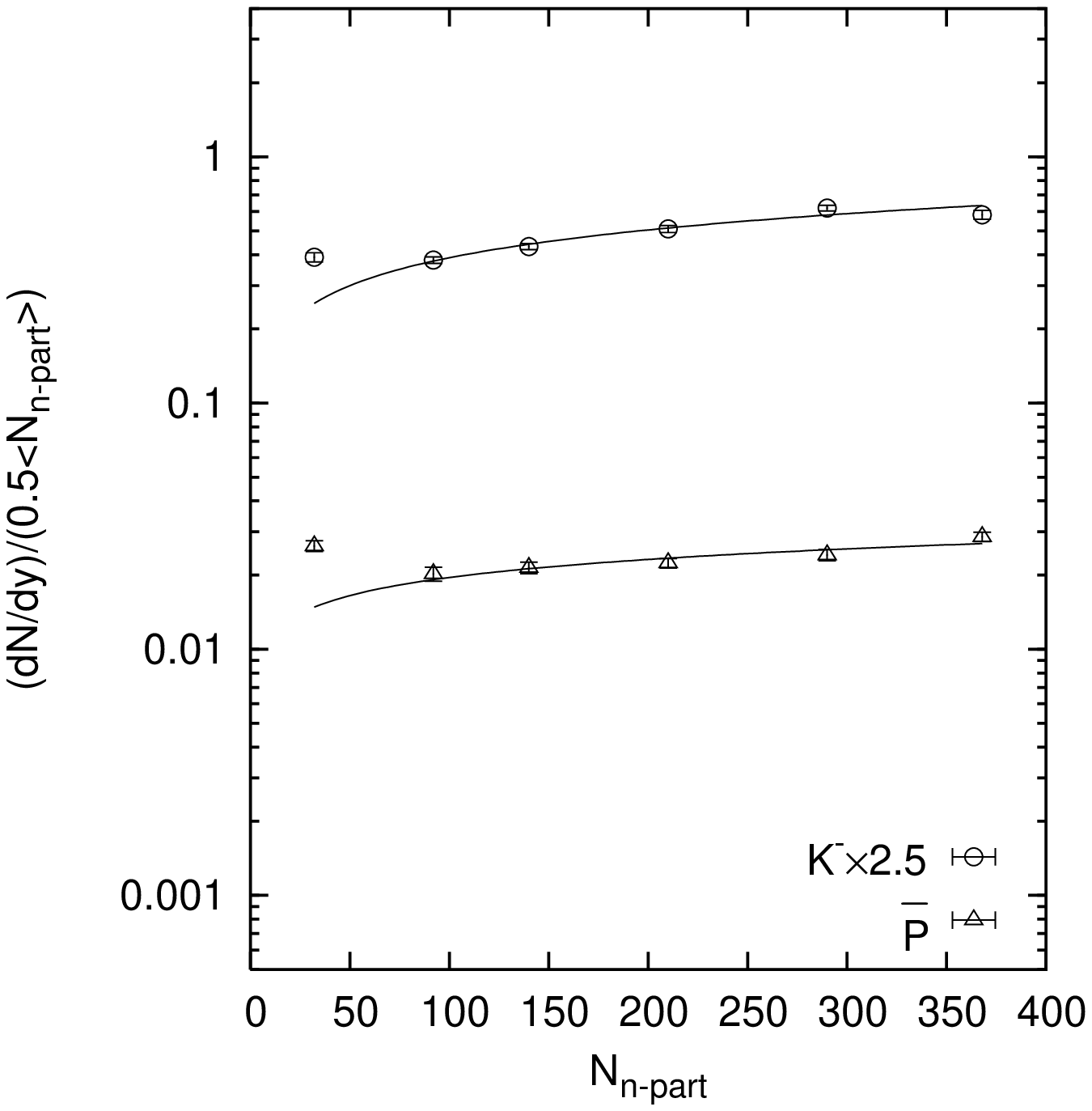}
  \end{minipage}}%
\vspace{.01cm}
 \subfigure[]{
  \begin{minipage}{.5\textwidth}
   \centering
\includegraphics[width=8cm]{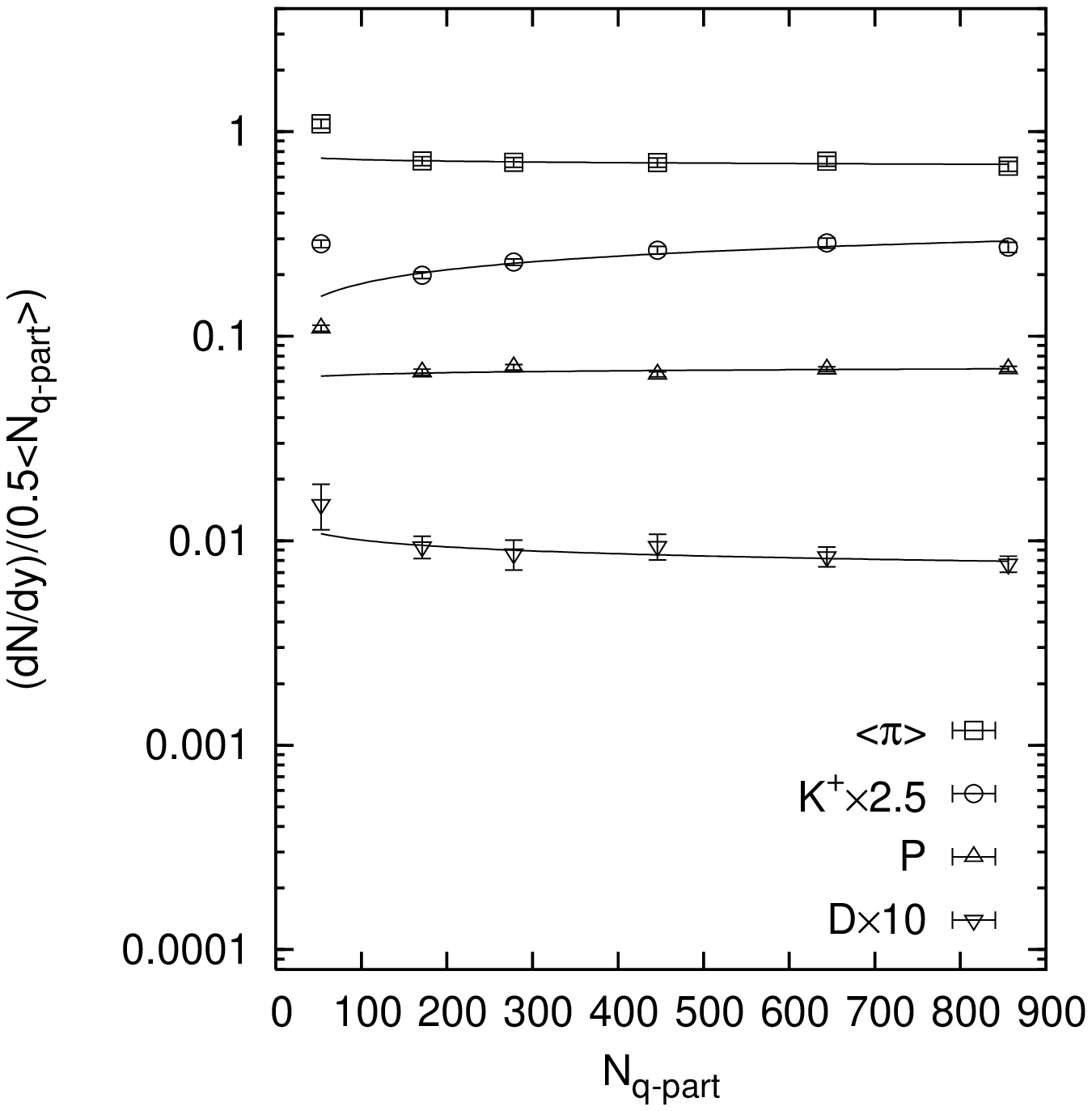}
  \end{minipage}}%
 \subfigure[]{
  \begin{minipage}{.5\textwidth}
   \centering
\includegraphics[width=8cm]{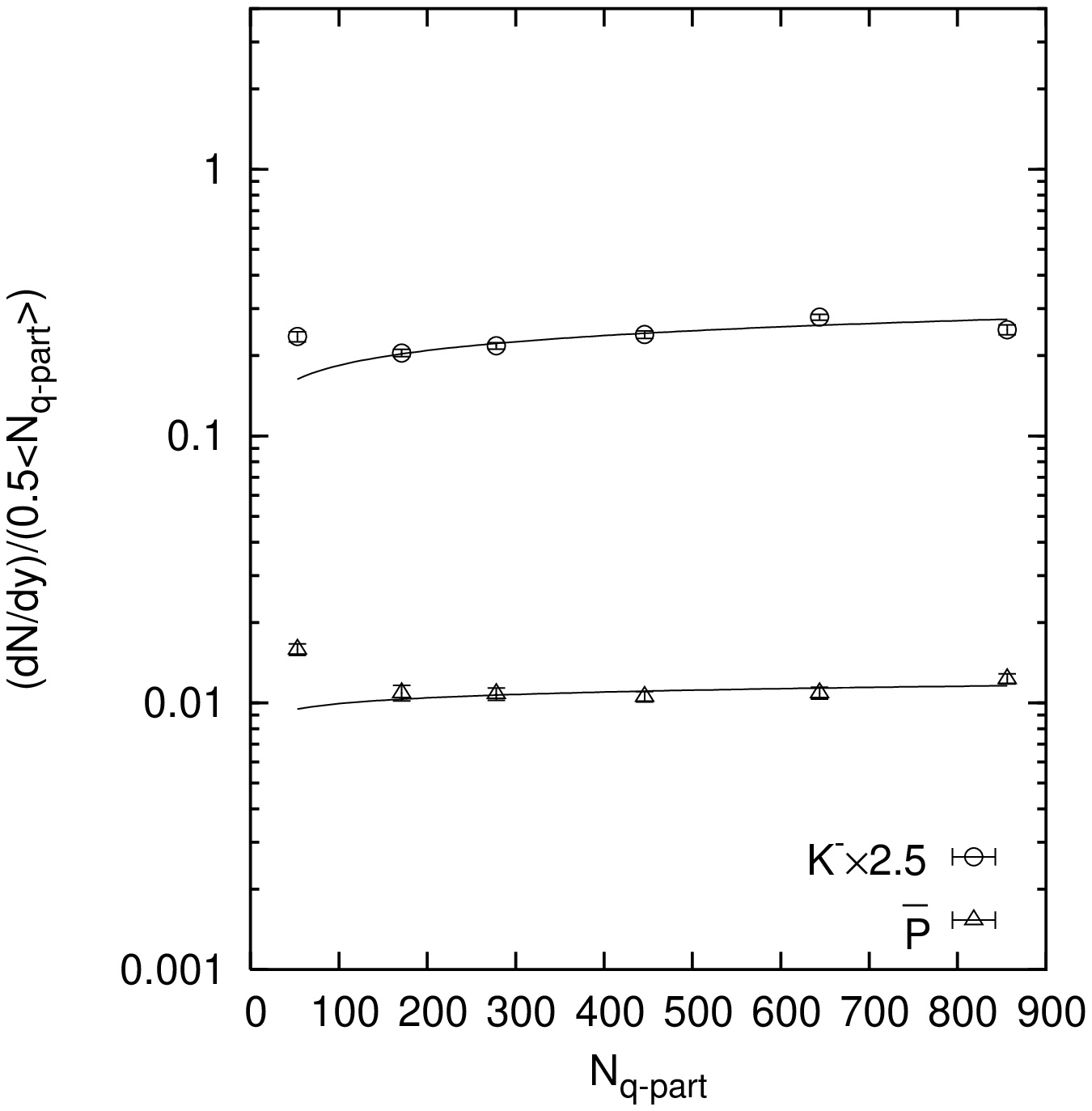}
  \end{minipage}}%
   \caption{Plots of integrated yields normalized by half of the  
   number of participant-nucleons[(a),(b)] or 
   constituent partons[(c),(d)] as a function of 
   centralities for production of the various identified secondaries in $Pb+Pb$ 
   collisions at $E_{Lab}=158A$ GeV\cite{Bachler1,Anticic1}. 
   The solid curves represent the fits obtained on the basis of 
   eqn.(5)[(a),(b)] and eqn.(6)[(c),(d)].}
\end{figure}

\begin{figure}
 \subfigure[]{
  \begin{minipage}{.5\textwidth}
   \centering
\includegraphics[width=8cm]{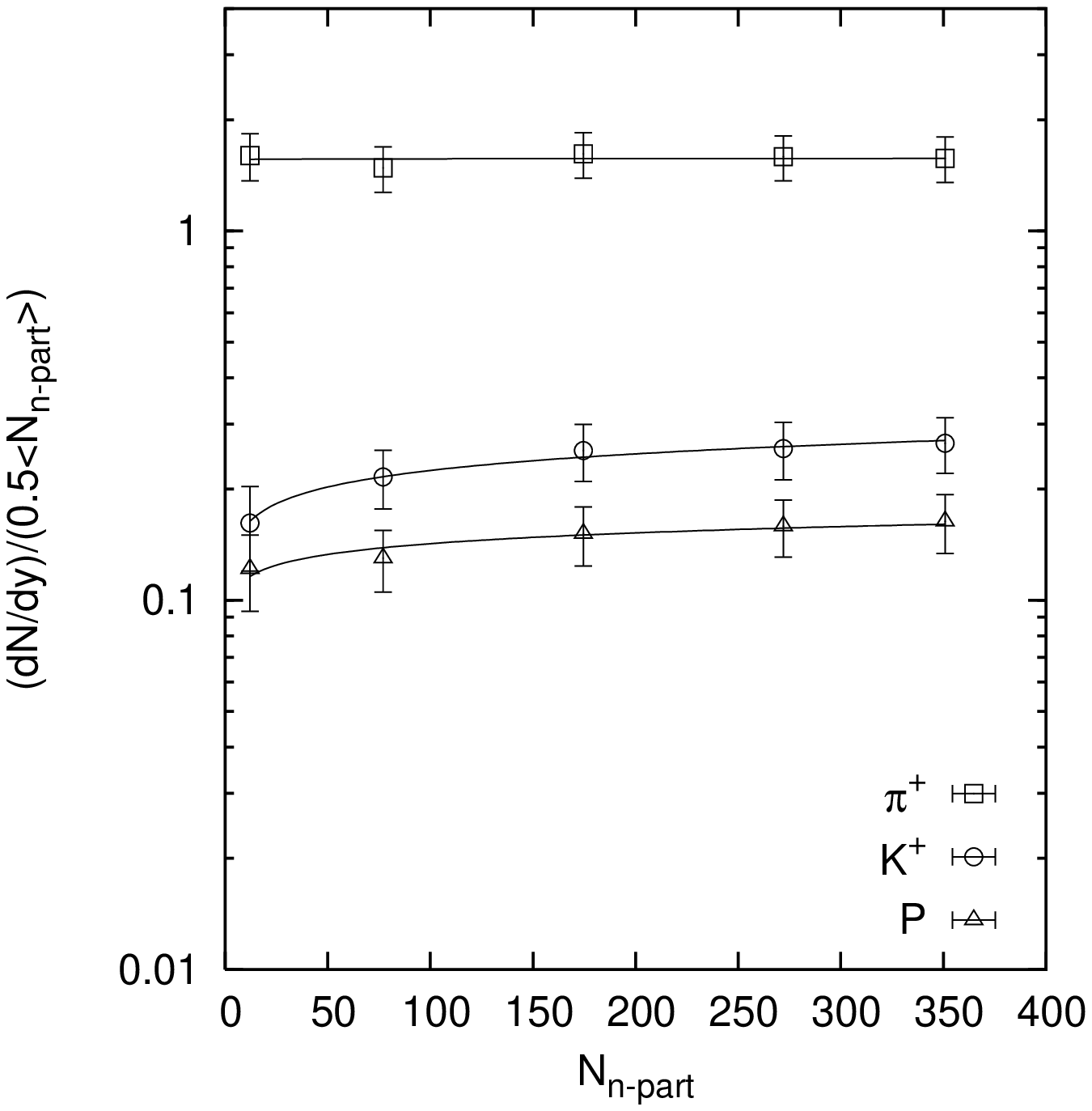}
  \end{minipage}}%
 \subfigure[]{
  \begin{minipage}{.5\textwidth}
   \centering
\includegraphics[width=8cm]{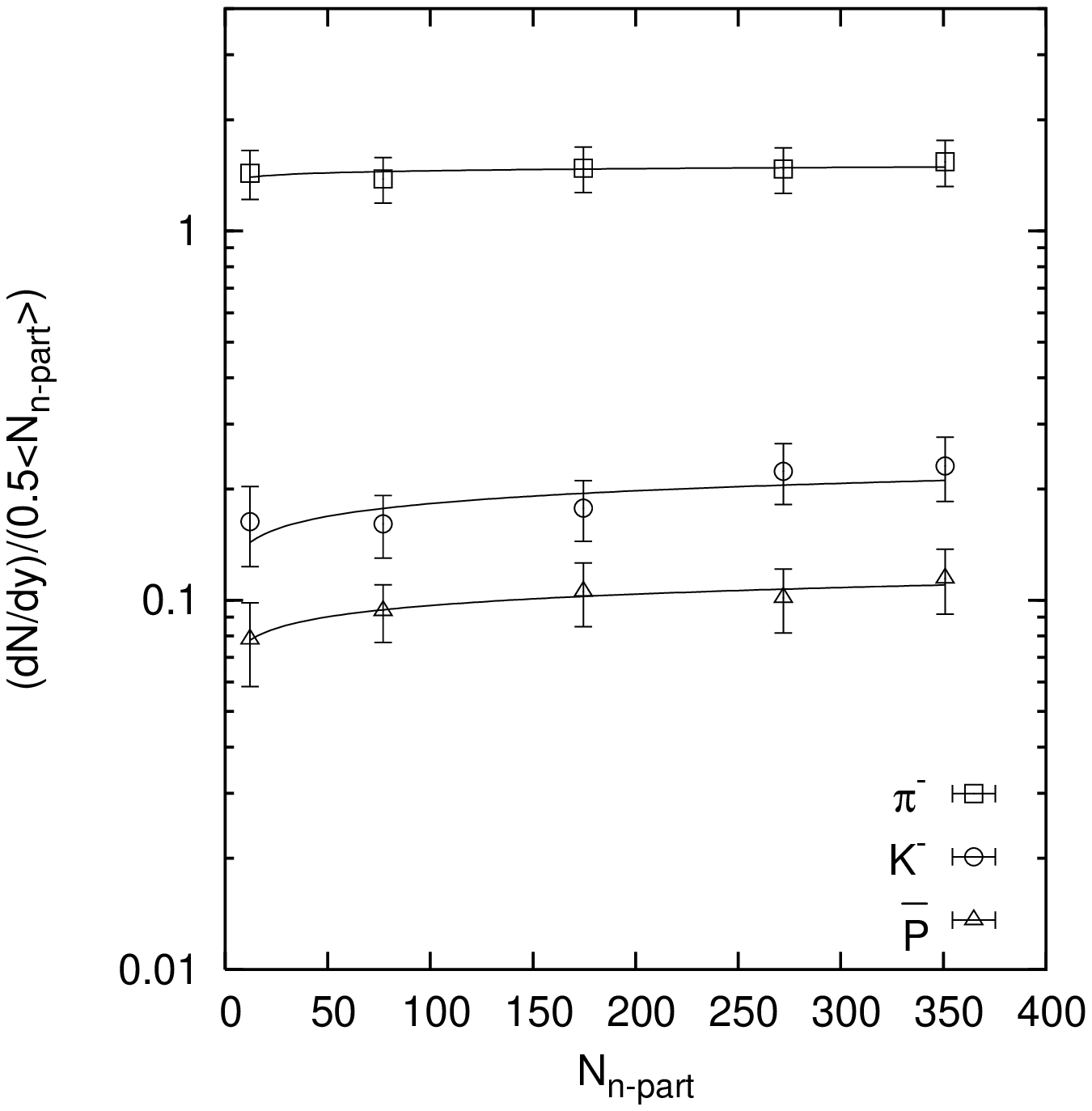}
  \end{minipage}}%
\vspace{.01cm}
 \subfigure[]{
  \begin{minipage}{.5\textwidth}
   \centering
\includegraphics[width=8cm]{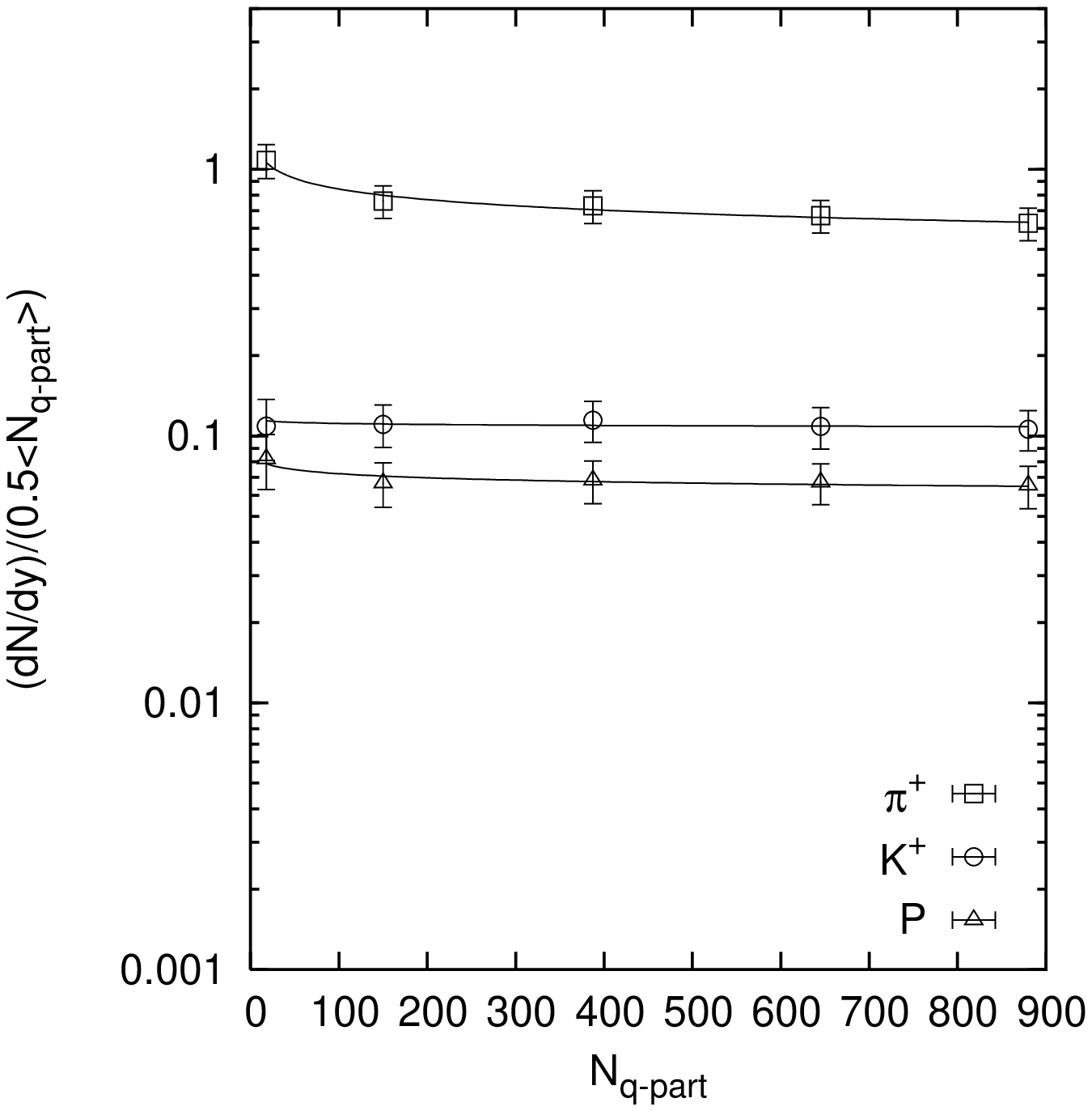}
  \end{minipage}}%
 \subfigure[]{
  \begin{minipage}{.5\textwidth}
   \centering
\includegraphics[width=8cm]{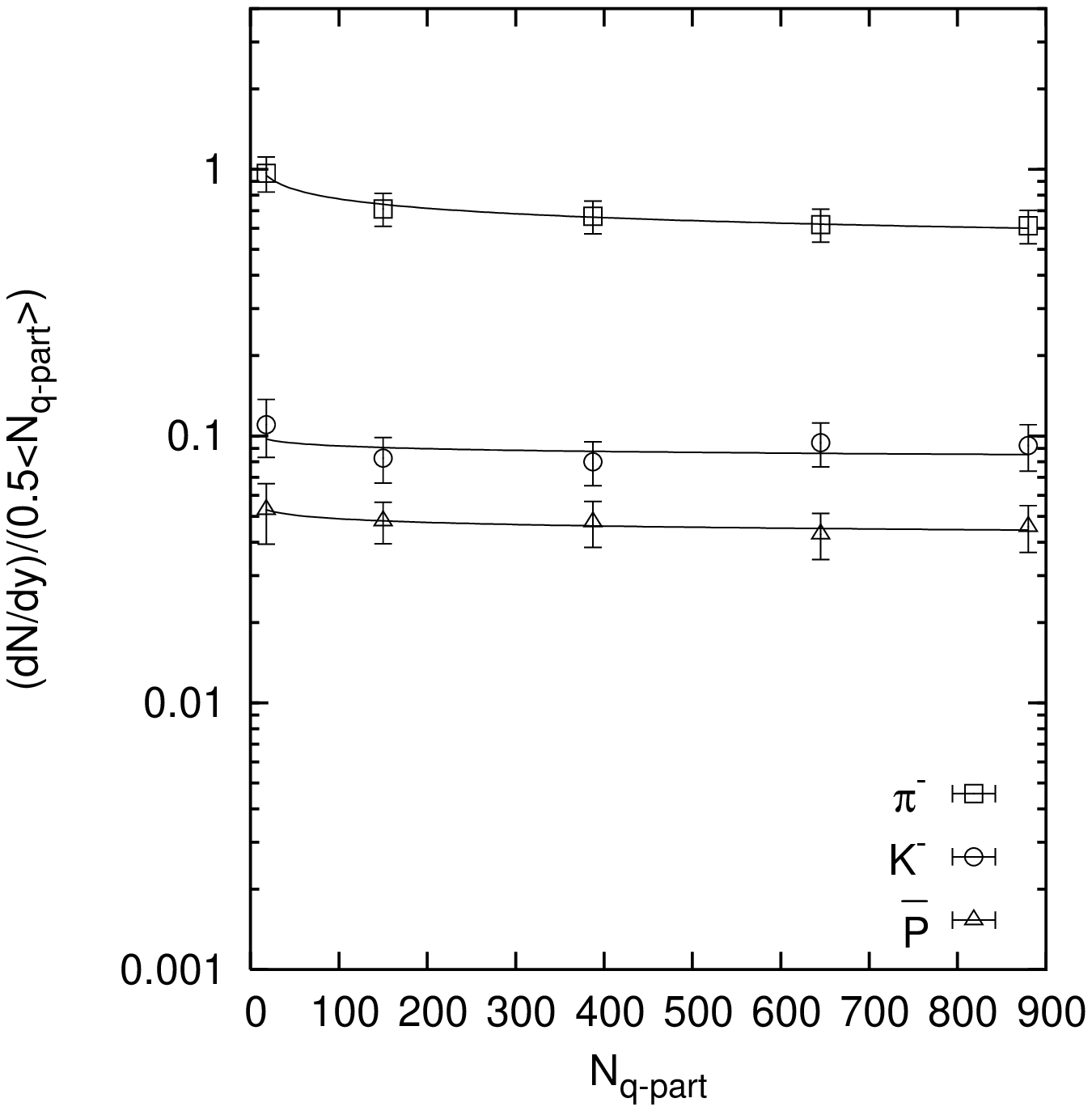}
  \end{minipage}}%
   \caption{Plots of integrated yields normalized by half of the  
   number of participant-nucleons[(a),(b)] or 
   constituent partons[(c),(d)] as a function of 
   centralities for production of the various identified secondaries in $Au+Au$ 
   collisions at $\sqrt{s_{NN}}=130$ GeV\cite{Adcox1}. The solid curves represent
   the fits obtained on the basis of eqn.(5)[(a),(b)] and eqn.(6)[(c),(d)].}
\end{figure}

\begin{figure}
 \subfigure[]{
  \begin{minipage}{.5\textwidth}
   \centering
\includegraphics[width=8cm]{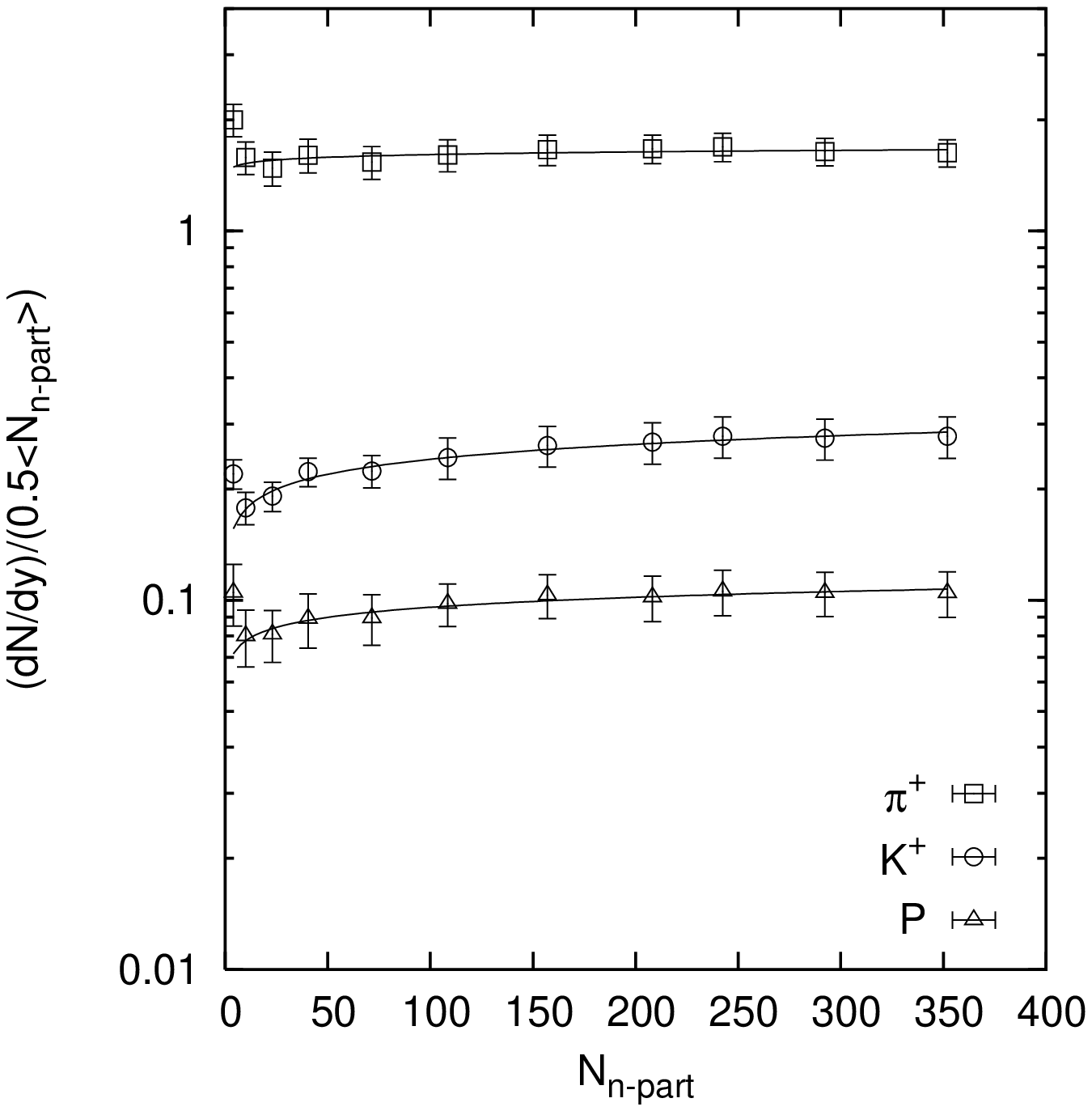}
  \end{minipage}}%
 \subfigure[]{
  \begin{minipage}{.5\textwidth}
   \centering
\includegraphics[width=8cm]{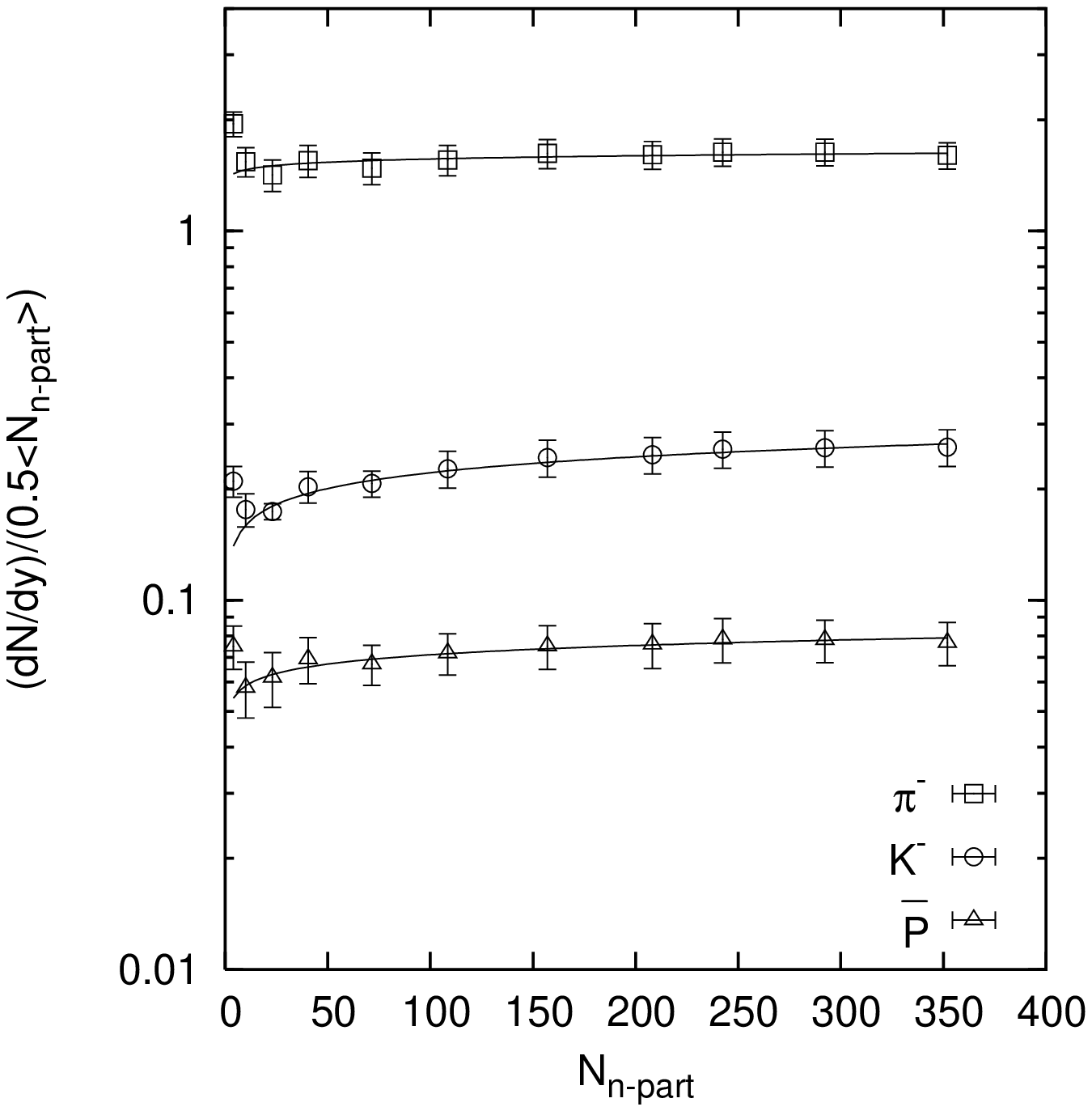}
  \end{minipage}}%
\vspace{.01cm}
 \subfigure[]{
  \begin{minipage}{.5\textwidth}
   \centering
\includegraphics[width=8cm]{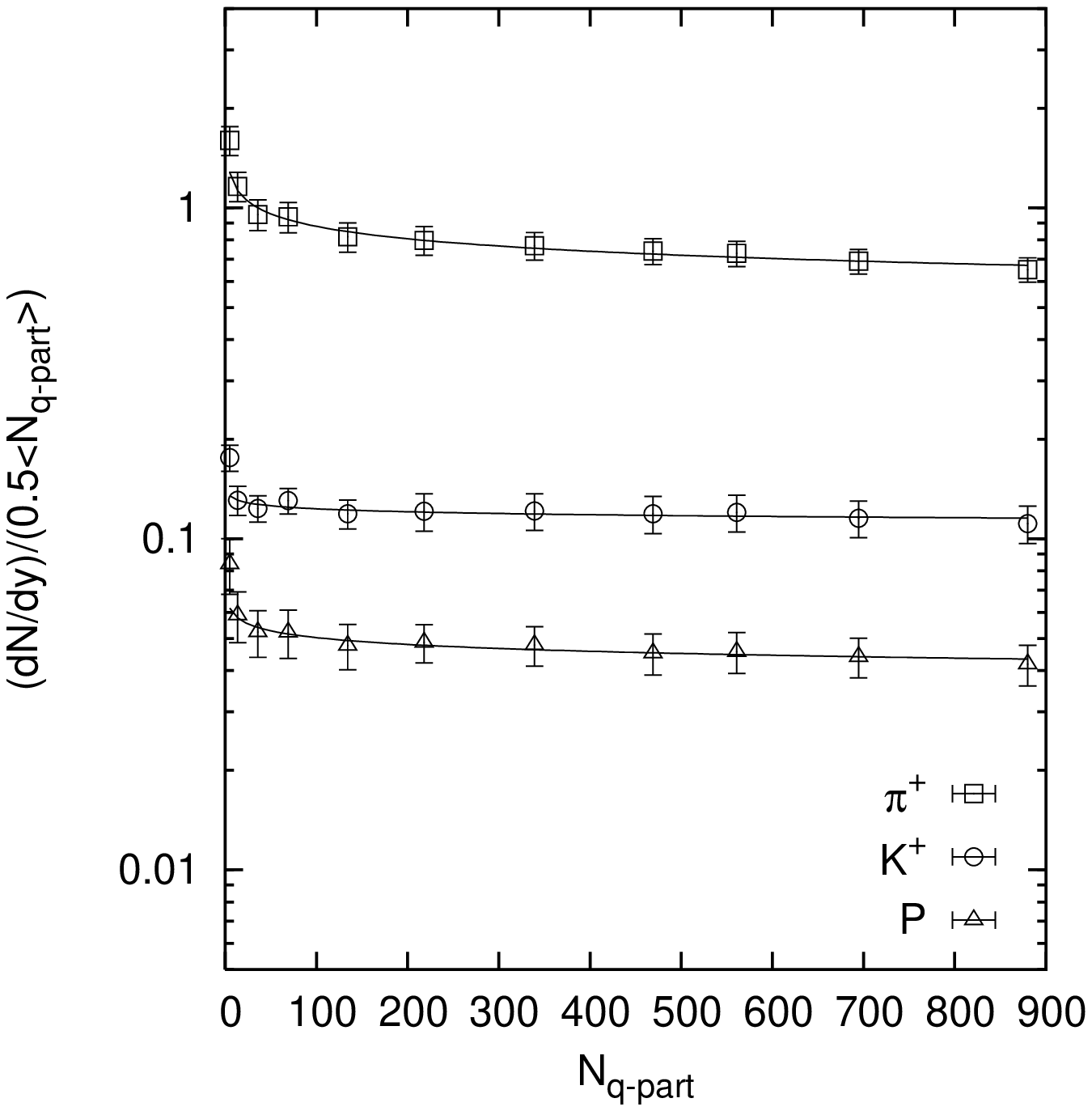}
  \end{minipage}}%
 \subfigure[]{
  \begin{minipage}{.5\textwidth}
   \centering
\includegraphics[width=8cm]{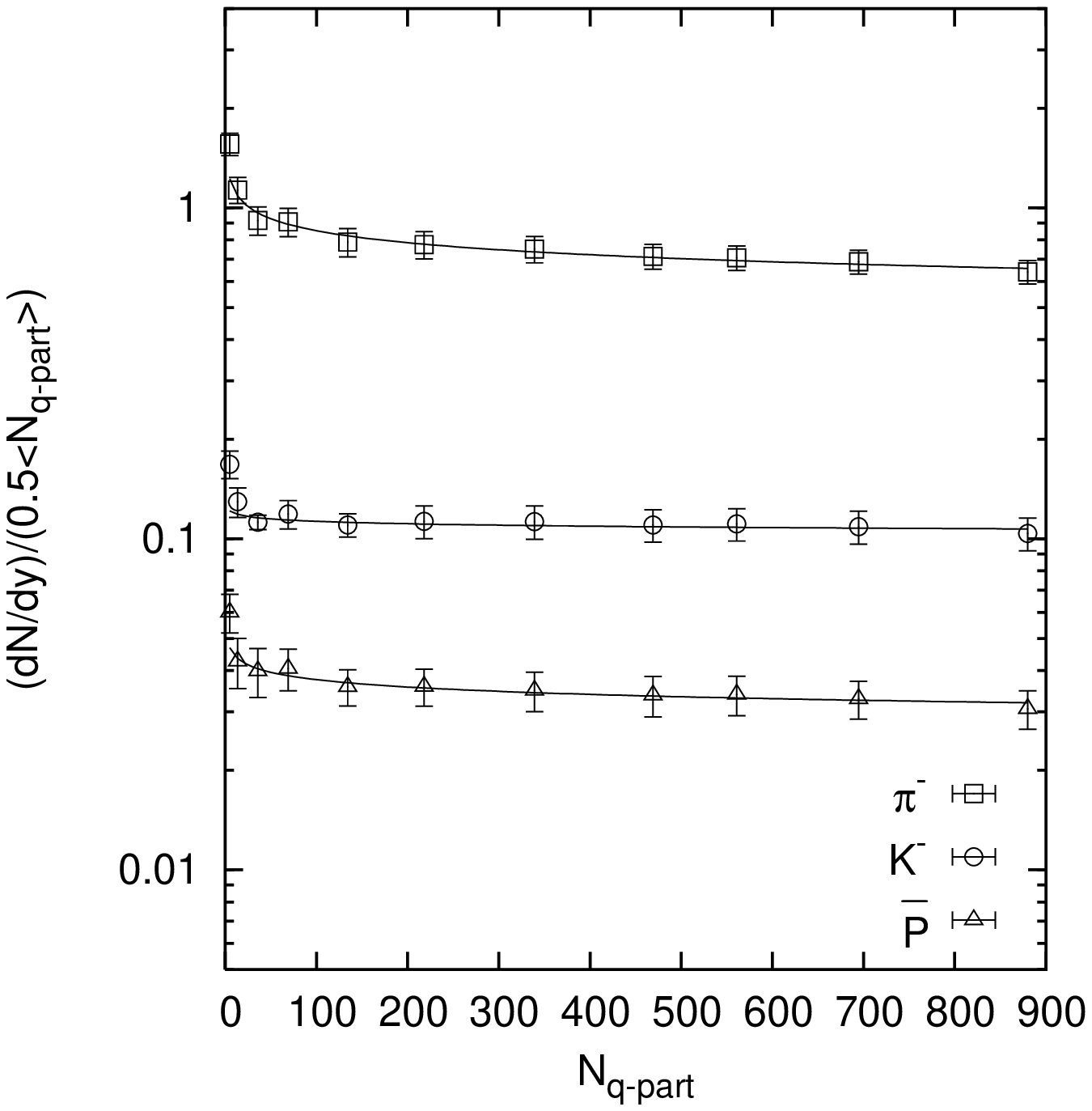}
  \end{minipage}}%
   \caption{Plots of integrated yields normalized by half of the  
   number of participant-nucleons[(a),(b)] or 
   constituent partons[(c),(d)] as a function of 
   centralities for production of the various identified secondaries in $Au+Au$ 
   collisions at $\sqrt{s_{NN}}=200$ GeV\cite{Adler1}. The solid curves represent
   the fits obtained on the basis of eqn.(5)[(a),(b)] and eqn.(6)[(c),(d)].}
\end{figure}

\begin{figure}
 \subfigure[]{
  \begin{minipage}{1\textwidth}
   \centering
\includegraphics[width=8cm]{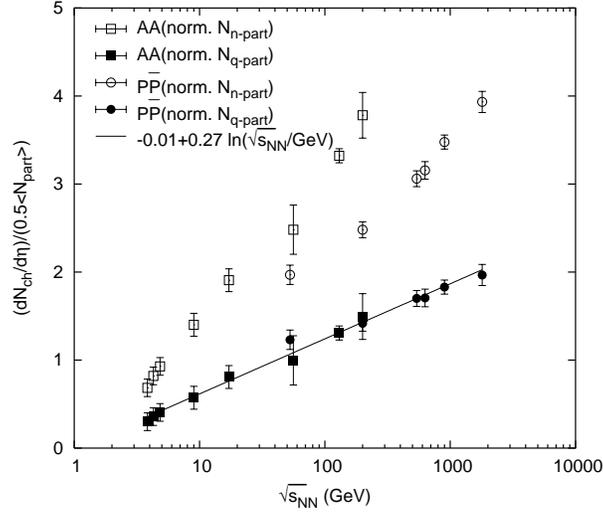}
  \end{minipage}}%
\vspace{.01cm}
 \subfigure[]{
  \begin{minipage}{1\textwidth}
   \centering
\includegraphics[width=8cm]{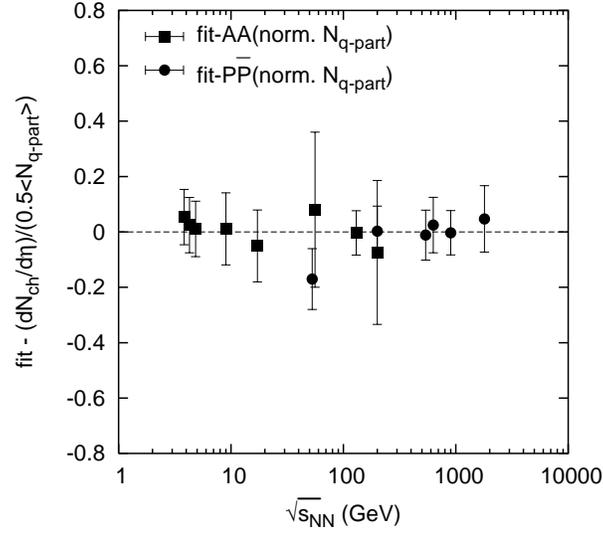}
  \end{minipage}}%
\caption{(a) Energy dependences of integrated yields for charged hadrons produced in 
different nucleus-nucleus($A+A$) collisions at AGS, SPS and RHIC energies and for 
the same produced in $P+\bar{P}$ collisions at ISR energies(Fig.3 of
Ref.\cite{Back1}). The open 
boxes and open circles provide the data when normalized by half of the average 
number of participant nucleons for nucleus-nucleus and $P\bar{P}$ collisions 
respectively. The solid boxes and solid circles show the same result for $A+A$ and
$P+\bar{P}$ collisions when normalized by half of the number of the participant-partons.
The solid line provide a fit[eqn.(7)] for the nucleus-nucleus and proton-antiproton data when
normalized by the half of the corresponding average number of constituent partons. (b)
The plot of closeness of the different sets of data with respect to the aforesaid fit.}
\end{figure}

\end{document}